\documentclass[sigconf,authorversion,screen]{acmart}

\usepackage[normalem]{ulem}
\usepackage{booktabs} 
\usepackage{color}
\usepackage{multirow}
\usepackage{subcaption}
\usepackage{caption}
\usepackage{enumitem}
\usepackage{hyperref}
\usepackage[multiple]{fnpct}
\usepackage{siunitx}
\usepackage{titlecaps}
\usepackage{colortbl}
\usepackage{color}

\usepackage{adjustbox}
\usepackage{array}
\newcolumntype{P}[1]{>{\centering\arraybackslash}p{#1}}

\usepackage{tikz}
\def\checkmark{\tikz\fill[scale=0.4](0,.35) -- (.25,0) -- (1,.7) -- (.25,.15) -- cycle;} 
\newcommand{\cross}{$\mathbin{\tikz [x=1.4ex,y=1.4ex,line width=.2ex] \draw (0,0) -- (1,1) (0,1) -- (1,0);}$}%

\newcommand{\myparagraph}[1]{\subsubsection*{\textbf{#1}}}

\newcommand{\studyone}{{\textit{answerability study}}}
\newcommand{\studytwo}{{\textit{viewpoints study}}}

\setcopyright{none}

\AtBeginDocument{%
  \providecommand\BibTeX{{%
    \normalfont B\kern-0.5em{\scshape i\kern-0.25em b}\kern-0.8em\TeX}}}

\copyrightyear{2024} 
\acmYear{2024} 
\setcopyright{acmlicensed}\acmConference[SIGIR-AP '24]{Proceedings of the 2024 Annual International ACM SIGIR Conference on Research and Development in Information Retrieval in the Asia Pacific Region}{December 9--12, 2024}{Tokyo, Japan}
\acmBooktitle{Proceedings of the 2024 Annual International ACM SIGIR Conference on Research and Development in Information Retrieval in the Asia Pacific Region (SIGIR-AP '24), December 9--12, 2024, Tokyo, Japan}
\acmDOI{10.1145/3673791.3698409}
\acmISBN{979-8-4007-0724-7/24/12}

\title[Can Users Detect Biases or Factual Errors in Generated Responses?]{Can Users Detect Biases or Factual Errors in Generated Responses in Conversational Information-Seeking?}

\author{Weronika Łajewska}
\orcid{0000-0003-2765-2394}
\affiliation{%
  \institution{University of Stavanger}
  \city{Stavanger}
  \country{Norway}
}
\email{weronika.lajewska@uis.no}

\author{Krisztian Balog}
\orcid{https://orcid.org/0000-0003-2762-721X}
\affiliation{%
  \institution{University of Stavanger}
  \city{Stavanger}
  \country{Norway}
}
\email{krisztian.balog@uis.no}

\author{Damiano Spina} 
\orcid{https://orcid.org/0000-0001-9913-433X}
\affiliation{
\institution{RMIT University}
  \city{Melbourne}
  \country{Australia}
}
\email{damiano.spina@rmit.edu.au}

\author{Johanne Trippas} 
\orcid{https://orcid.org/0000-0002-7801-0239}
\affiliation{
\institution{RMIT University}
  \city{Melbourne}
  \country{Australia}
}
\email{j.trippas@rmit.edu.au}

\begin{abstract}
Information-seeking dialogues span a wide range of questions, from simple factoid to complex queries that require exploring multiple facets and viewpoints. When performing exploratory searches in unfamiliar domains, users may lack background knowledge and struggle to verify the system-provided information, making them vulnerable to misinformation. We investigate the limitations of response generation in conversational information-seeking systems, highlighting potential inaccuracies, pitfalls, and biases in the responses. The study addresses the problem of \emph{query answerability} and the challenge of \emph{response incompleteness}. 
Our user studies explore how these issues impact user experience, focusing on users' ability to identify biased, incorrect, or incomplete responses. We design two crowdsourcing tasks to assess user experience with different system response variants, highlighting critical issues to be addressed in future conversational information-seeking research. 
Our analysis reveals that it is easier for users to detect response incompleteness than query answerability and user satisfaction is mostly associated with response diversity, not factual correctness. \emph{This is a preprint version of the paper published at SIGIR-AP'24, including appendices with further analysis of our experiments.}

\end{abstract}



\begin{document}
\maketitle

\section{Introduction}

Conversational information-seeking (CIS) interactions enable users to fulfill complex information needs, navigate unknown domains, ask follow-up questions, and provide feedback via a series of natural language dialogues~\citep{Zamani:2023:FNT}.
CIS research currently centers on retrieval components, such as passage retrieval, re-ranking, and query rewriting~\cite{Zamani:2023:FNT,Owoicho:2022:TRECb}. However, the core difficulty lies in effectively assembling the retrieved information into a trustworthy and reliable conversational response that the user will ultimately interact with. The task of synthesizing information from the top retrieved passages into a single response is called \emph{conversational response generation}~\citep{Ren:2021:TOIS}. Unfortunately, responses generated by CIS systems are susceptible to limitations, including hallucinations when no answer is found
~\citep{Ji:2023:ACMa}, biased responses that only partially answer the question~\citep{Gao:2020:Information}, or factual error presentation~\citep{Tang:2023:ACL}. 
These limitations potentially lead to inaccuracies, pitfalls, and biases, which may not always be evident to users, particularly those who lack familiarity with the search topic or the necessary background knowledge.
As individuals without specific training can only distinguish between human-generated and auto-generated texts at a level close to random chance~\citep{Clark:2021:ACL-IJNLP}, factually incorrect, unsupported, biased, or incomplete information may be easily overlooked.

This paper investigates users' ability to recognize pitfalls in CIS systems related to \emph{query answerability} and \emph{response incompleteness} (see~Table~\ref{tab:problems_examples}). We hypothesize that untrained users cannot identify these problems in CIS interactions. More specifically, we aim to address the following research questions: 
\begin{itemize}[leftmargin=0.8cm,labelsep=0.1cm]
    \item[\textbf{RQ1:}] Can users effectively recognize problems related to \emph{query answerability} and \emph{response incompleteness} in system responses?
    \item[\textbf{RQ2:}] How do factually incorrect, inaccurate, incomplete, and/or biased responses impact the user experience?
\end{itemize}
We design and conduct two crowdsourcing-based studies to determine whether users can effectively recognize these two problems in responses based on a subset of topics from the TREC Conversational Assistance (CAsT) datasets~\cite{Dalton:2020:TREC, Owoicho:2022:TRECb} with manually injected inaccuracies or biases in a controlled manner. 
Query answerability can be defined at different levels, which includes determining whether answer is present within the top relevant passages, the entire corpus, or general world knowledge. Additionally, when ``no answer found'' is the outcome, the system must transparently reveal this to the user and suggest ways to continue the conversation. In this paper, we focus on (i) the consequences of generating response from passages that do not contain the answer, which result in non-factual or hallucinated content, and (ii) the impact of source presentation.
The variants of responses in the \studyone{}~(i.e., study one) differ in factual correctness~\citep{Kryscinski:2020:EMNLP} and the presence/validity of the information source~\citep{Bolotova-Baranova:2023:ACL,Liu:2023:EMNLP}. 
The issue of response incompleteness encompasses a range of challenges, such as presenting biased information that covers only one facet or viewpoint, determining which pieces of information to include given response length limitations, and transparency regarding the relevant information not covered. In this paper, we focus on the subtask of viewpoint/facet diversification and examine the impact of balanced viewpoint coverage in responses.
The variants of the responses in the \studytwo{}~(i.e., study two) vary in diversity (in terms of viewpoints and/or facets)~\citep{Helberger:2018:Information} and balance in covering various viewpoints/facets in the response.

Results of the \studyone{} show that users cannot recognize factual errors in system responses. Additionally, a lack of source or an invalid source does not decrease their confidence in the response. On the other hand, according to the \studytwo{}, it is easier for users to identify problems with viewpoint diversity and balance, and recognize if the response is biased or incomplete (\textbf{RQ1}).
Moreover, the satisfaction ratings and comments from our user experience questionnaire reveal that overall system response satisfaction is associated with the investigated response dimensions 
(i.e., factual correctness and source presence/validity for the \studyone{}; diversity and balanced viewpoints/facets presentation for the \studytwo{}), even though the satisfaction and response dimension ratings do not fully align with the comments (\textbf{RQ2}).
The findings from our two studies reveal  that the fluency of CIS responses may compromise users' capability to spot inaccurate information---even in responses attributed to the source.

In summary, the main contributions of this work include: 
(1) a novel methodology to study how users perceive query answerability and response incompleteness in CIS, 
(2) a manually curated dataset of CIS responses varying across multiple dimensions in terms of quality, (3) quantitative and qualitative analysis of how users recognize and perceive limitations in CIS responses. The developed resources (manually generated CIS responses, user studies results, data analysis scripts) and additional analysis are available at:
\url{https://github.com/iai-group/sigirap2024-resgen}.

\begin{table}[tp]
    \centering
    \footnotesize
    \captionsetup{skip=5pt} 
    \caption{Example problems of \emph{query answerability} and \emph{response incompleteness}: the first response contains factual errors and is based on sources that do not provide an answer to the question (Malbec wine is not produced in Penedès, Spain). The second response mentions multiple viewpoints, but only one is covered in detail, resulting in a biased answer.}
    \label{tab:problems_examples}
    \begin{tabular}{p{3.2cm}p{4.5cm}}
    \toprule
        \emph{Query Answerability} & \emph{Response Incompleteness} \\
        \midrule
        To combine hiking and Malbec wine, plan a trip to the Penedès region in Catalonia. You can explore the Montserrat mountain range, which offers fantastic hiking opportunities, and then visit renowned wineries in the Penedès, known for its exceptional Malbec wine production\ldots
        \newline
        \url{https://www.winetourism.com/wine-tasting-tours-in-penedes/} & The Watergate scandal had a profoundly negative impact on President Nixon's legacy, overshadowing many of his domestic achievements. It tarnished his reputation as a corrupt politician, making him a symbol of political scandal and misconduct in both American politics and popular culture. While he did enact significant legislation like creating the Environmental Protection Agency, his presidency is primarily remembered for the scandal, leading to his resignation and becoming synonymous with political corruption and disgrace. \\
        \bottomrule
    \end{tabular}
    \vspace{-\baselineskip}
\end{table}

\section{Related Work}
\label{sec:related}

In search engine results pages (SERPs), users are presented with a ranked list of results, each accompanied by a source, providing a broader context for the requested information.
In contrast, response generation in CIS systems offers limited information in the form of natural language responses, typically consisting of several sentences~\citep{Ren:2021:TOIS}. This shift has been facilitated by the widespread adoption of generative language models, which enable systems to produce fluent and coherent responses~\citep{Zhang:2020:ACL}. 
One of the primary expectations from generated responses is to equip users with the necessary tools for assessing the reliability and accuracy of the provided information~\citep{Liu:2023:CHIIR}. 
User studies on CIS human-system interactions highlight desired response features, such as opinion aggregation, including information sources to ensure credibility, response verifiability, as well as balance between direct answers and expanded information to address the conciseness/completeness tradeoff~\citep{Vtyurina:2017:CHI, Trippas:2020:Information, Liu:2023:EMNLP}.
CIS systems aim to provide users with responses that encapsulate relevant information from multiple documents, creating a more natural and user-friendly experience~\citep{Culpepper:2018:SIGIR,Zamani:2023:FNT}. 
Multi-document summary generation has been studied in the context of non-factoid question answering~\citep{Bolotova-Baranova:2023:ACL}, generative search engine results verifiability~\citep{Liu:2023:EMNLP}, and generating summaries from retrieved results that was one of the tasks proposed in TREC CAsT'22~\citep{Owoicho:2022:TRECb}.

However, transitioning from SERPs to conversational responses presents distinct challenges \cite{Wu:2003:J.,Wu:2005:Information,Trippas:2018:CHIIRb, Schuster:2023:arXiv}. Even though the ranking of top retrieved passages should ensure fairness~\citep{Gao:2020:Information} and viewpoint diversification~\citep{Gao:2020:Information, Draws:2021:SIGIR, Sakaeda:2022:NAACL-HLT}, it is a non-trivial task to synthesize those passages into a reliable, trustworthy, and concise response. 
Additionally, cognitive biases, such as anchoring bias and confirmation bias, can impact user interactions with the search system, potentially disrupting the overall user experience~\citep{Azzopardi:2021:CHIIR, Liu:2023:CHIIR, White:2014:Ja, McKay:2022:CHIIR,Kiesel:2021:CUI,Spina:2024:RMIT_report}. 
Response generation from retrieved passages faces additional challenges related to temporal considerations~\citep{Campos:2015:ACM}, biased queries~\citep{Azzopardi:2021:CHIIR, Kiesel:2021:CUI}, source subjectivity, unanswerability~\citep{Choi:2018:EMNLP, Rajpurkar:2018:ACL, Reddy:2019:TACL,PathiyanCherumanal:2024:CHIIR}, the lack of expert knowledge
, and additional issues related to text aggregation that may introduce hallucinations and factual errors~\citep{Tang:2023:ACL}.
Given the potential flaws that may result from these challenges, 
conversational response generation 
should involve system revealment and promote a more informed user experience \citep{Azzopardi:2018:CAIRc, Radlinski:2017:CHIIRd, Lajewska:2024:SIGIR}. In the proposed user studies, we aim to investigate to what extent users are unaware of the inaccuracies in the responses, and what the scale of the problem is in CIS. Nevertheless, providing users with an understanding of the search space and transparently conveying the system's certainty and potential pitfalls are essential for promoting user trust and informed interactions with the system.

Evaluating response quality in CIS systems presents unique challenges, as traditional offline evaluation measures like ROUGE~\citep{Lin:2004:ACL} (commonly used for evaluating summaries) and  NDCG~\citep{Jarvelin:2002:ACM} (for evaluating passage rankings) 
fail to fully capture the complexities of conversational context, multi-turn dialogue coherence, and the overall user experience in conversational interactions.
Evaluating CIS responses from a user perspective involves multiple dimensions~\cite{Sakai:2023:arXiva}, including trust and fairness~\citep{Zamani:2023:FNT}, credibility~\citep{Bink:2022:CHIIR}, reliability~\citep{Lu:2021:CHI, Rechkemmer:2022:CHI}, verifiability~\citep{Liu:2023:EMNLP}, factual correctness, transparency (e.g., information sources, ranking, and consolidation process)~\citep{Shah:2022:CHIIR}, relevance, naturalness, conciseness~\citep{Owoicho:2022:TRECb}, informativeness (supporting user in increasing their information literacy)~\citep{Shah:2022:CHIIR}, perceived satisfaction, and usefulness~\citep{Hoeve:2020:arXiv, Zheng:2022:CHI, Cambazoglu:2021:CHIIR}. 
However, directly asking users to report on these metrics may not be reliable as users may interpret the concepts differently (the problem of indirect observables)~\citep{Kelly:2007:FNTa}.
To tackle this challenge, our research focuses on understanding the response dimensions that are (1) associated with user satisfaction and (2) affected by answerability and incompleteness issues.
\section{Methodology}
\label{sec:Methodology}

We aim to investigate if users can recognize inaccuracies in CIS system responses and how these inaccuracies impact the user experience---hereafter, we use \textit{response} to refer to \textit{CIS system response}. We conduct two crowdsourcing studies\footnote{This investigation was compliant with the ethics approval process of our institution.} employing a within-subject design that investigate the problems of: 
\begin{itemize}[leftmargin=*]
    \item Query answerability through an \studyone{} with the focus on factual errors and quality of the information sources accompanying the response. 
    \item Response incompleteness through a \studytwo{} with the focus on balance of viewpoints and/or facets  in the response.
\end{itemize}
\noindent
For each study, we select ten queries susceptible to one of the identified problems (i.e., answerability or incompleteness).
For each query, we manually create response variants differing in terms of two controlled dimensions
(1) factual correctness and 
(2) source presence/validity in the \studyone{}; and 
(1) facet/viewpoint diversity and 
(2) balanced facet/viewpoint presentation in the \studytwo{}. 
Workers are presented with a set of queries with responses and asked to indicate their perception of the controlled dimensions listed above, as well as their overall satisfaction. We consider a simplified scenario involving a set of topics that are particularly susceptible to these issues, and we manually introduce isolated, easily detectable errors. We acknowledge that in real-world conversations such errors are likely to be much harder to identify. This paper presents only a preliminary study, and exploring more realistic and complex scenarios is left for future work.

We aim to investigate users' ability to detect pitfalls in responses in a scenario that closely mirrors real-life system interactions. In actual situations, a user poses a query, receives a single system response, and must then judge whether this response is useful and satisfying. 
To replicate this setting, we provide each worker with a set of identical queries and a single version of the response for each query. This way, we may include different variants of the response in one task without the differences being too conspicuous when all possible variants of the response for a given query are presented consecutively. 
These response sets are carefully balanced in terms of accuracy, ensuring that users encounter in their microtasks---hereafter, Human Intelligence Tasks (HITs)---responses of different quality, without those differences being overly apparent.

\subsection{Experimental Design}
\label{subsec:procedure}
Crowd workers are presented with ten query-response pairs in each HIT and asked to assess the provided responses. 
Responses differ in their quality and accuracy along different controlled dimensions. Each response is an instance of one of the experimental conditions.
In the \studyone{}, we consider four different experimental conditions \(EC^A\) (resulting in four response variants for each query), and in the \studytwo{}, three \(EC^V\) (with three response variants for each query). 
The experimental conditions of manually crafted responses for both user studies are described in~Section~\ref{subsubsec:responses}.

Both our studies follow the Graeco-Latin square design, which ensures the rotation and randomization of queries and response variants, as well as no overlap in sets of query-response pairs between HITs~\citep{Kelly:2007:FNTa}.
Each query-response pair appears in three different HITs, where each HIT contains a different set of ten query-response pairs. Query-response pairs appear in the HITs in a random order.
Considering grouping factors that arise whenever one annotator rates multiple responses, we ensure that each crowd worker completed only a single HIT for a given user study (but they were allowed to participate in both user studies). This way, we attempt to balance the need for a large enough annotator pool with a sufficient task size to be worthwhile to the crowd workers~\citep{Steen:2021:ACL}.

\subsection{Tasks}
\label{subsec:tasks}

\begin{figure}
    \centering
    \includegraphics[width=0.25\textwidth]{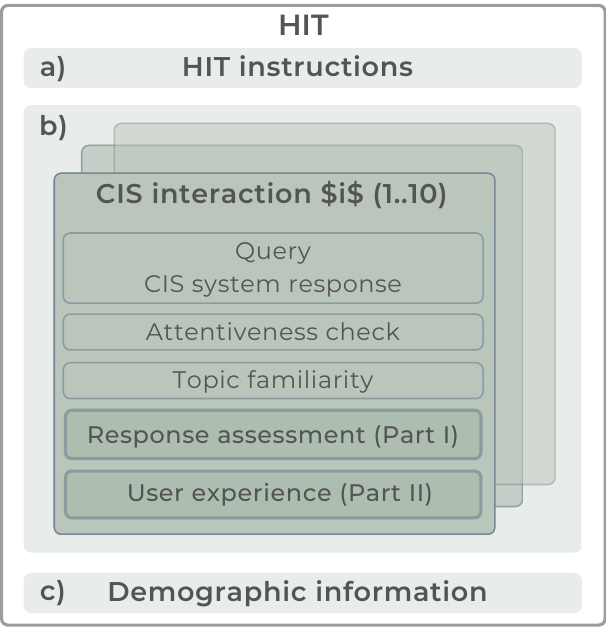}
    \captionsetup{aboveskip=5pt}
    \caption{High-level design of the user studies.}
    \label{fig:study_design}
\end{figure}

\begin{figure*}[tp]
    \centering
    \includegraphics[width=1.0\textwidth]{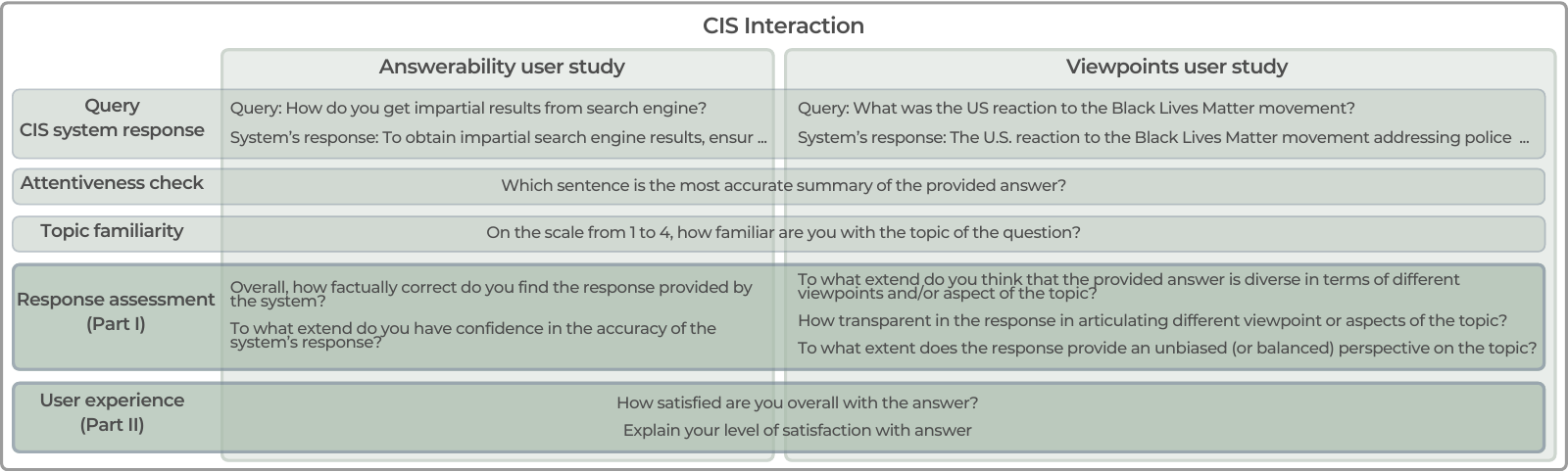}
    \captionsetup{aboveskip=5pt}
    \caption{Questions provided to crowd workers in our user studies.
    }
    \label{fig:study_questions}
\end{figure*}

The design of the \studyone{} and the \studytwo{} follows the same principle: workers are asked to complete one HIT, consisting of ten query-response pairs. The task consists of 
(a) HIT instructions;
(b) ten CIS interactions; and 
(c) demographics questionnaire as seen in~Figure~\ref{fig:study_design}. 
Workers are not given specific examples of query-response pairs in the instructions to avoid biasing them.
We decompose each user study into multiple subsections using independent CIS interactions to facilitate atomic microtask crowdsourcing~\citep{Gadiraju:2015:IEEE}.
Each CIS interaction contains one query-response pair, followed by (1) a corresponding attentiveness check, (2) a measurement of the worker's familiarity with the topic, (3) a CIS response assessment (Part I)
, and (4) a measurement of user experience (Part II).\footnote{The only difference between the \studyone{} and \studytwo{} are the response dimensions for which we are collecting crowd workers' ratings in Part I.}
The wording of the questions in all parts of the user studies follows questions proposed by~\citet{Tang:2022:NAACL-HLT} for evaluating the factual consistency of summaries (see~Figure~\ref{fig:study_questions}). 
Both studies finish with a short demographics questionnaire asking workers' age, education level, and gender. 

\subsubsection{\textbf{Attentiveness Check}}
\label{subsubsec:Attentiveness check}
We present workers with an additional question for each CIS interaction for which we have a ground truth answer to serve as an attention check, 
which enables us to detect poorly performing workers, cheat submissions, or bots~\citep{Gadiraju:2015:IEEE}. 
Each attention check question consists of three sentences related to the query's topic, one of them being a summary of the provided response. Sentences are provided in a random order and workers are asked to select the best summary~\citep{Bolotova-Baranova:2023:ACL}. 
Submissions that failed on more than 3/10 attentiveness questions were rejected. 

\subsubsection{\textbf{Topic Familiarity}}
\label{subsubsec:Pre-task questionnaire}
In this part of the CIS interaction task, crowd workers are asked to rate their familiarity with the query topic to help us assess the task difficulty and condition the collected data on users' background knowledge~\citep{Krishna:2021:NAACL-HLT}. 

\subsubsection{\textbf{Part I: Response Assessment}}
\label{subsubsec:CIS response assessment}
In Part I, workers are asked to evaluate the dimensions of the response presented for a given query. Since we are investigating different response dimensions
for \studyone{} and \studytwo{}, each study's response assessment part is different. The questions asked per study are related to the dimensions we identified for each problem and are answered by workers on four-point Likert scales.
To increase the ecological validity of our experiments (and avoid making the assessment task too artificial), the dimensions used to control the generation of response (\emph{controlled response dimensions}) do not always directly map to the dimensions that workers are asked to assess (\emph{user-judged response dimensions}) (see~Table~\ref{tab:answer_dimensions}). In the case of response dimension (2) in the \studyone{} (source presence/validity), simply asking workers whether the source is present or the link is valid would be too apparent and would violate the user study by directly suggesting some specific user behavior (i.e., clicking the link). Therefore, we attempt to capture this dimension by asking about the worker's confidence in the accuracy of the answer. In the case of response dimension (1) in the \studytwo{} (diversity), it is not enough to ask how diverse the topic is, since recognizing the lack of diversity requires some knowledge about the topic. Therefore, we include an additional user-judged response dimension related to transparency in articulating different viewpoints or facets of the topic. Dimension (2) in the \studytwo{} (balance) is provided with an additional explanation to ensure a common understanding of the underlying concept. Namely, we ask to assess the unbiased (or balanced) perspective on the topic.

\begin{table}[tp]
    \centering
    \footnotesize
    \captionsetup{skip=5pt} 
    \caption{Controlled vs. user-judged response dimensions.}
        \label{tab:answer_dimensions}
    \begin{tabular}{lcc}
        \toprule
         \multirow{2}{*}{\textbf{User Study}} & \multicolumn{2}{c}{\textbf{Response Dimension}} \\
        \cmidrule{2-3}
         & \textbf{Controlled} & \textbf{User-judged} \\
        \midrule
        \multirow{2}{*}{Answerability} & (1) Factual Correctness & Factual Correctness \\
         & (2) Source Presence/Validity & Confidence in Answer Accuracy \\
        \midrule
        \multirow{2}{*}{Viewpoint} & (1) Diversity & Diversity + Transparency \\
         & (2) Balance & Balance/Bias \\
        \bottomrule
    \end{tabular}
\end{table}

\subsubsection{\textbf{Part II: User Experience}}
\label{subsubsec:User experience}
In the final part of each CIS interaction, we pose a question about the overall satisfaction with the response (a proxy for the user experience).
It is followed by a required open text field for workers to elaborate on their decision. 

\subsection{Data Analysis Methods}
\label{subsec:data_analysis_methods}

To address RQ1, we assess if workers can detect flaws and inaccuracies in the responses based on their ratings for user-judged response dimensions.
We use two-way ANOVA~\citep{Kelly:2007:FNTa} for analyzing the results, where the different controlled response dimensions, representing different variants of the responses, are factor variables. A separate ANOVA is performed for each of the user-judged response dimensions (dependent variables) with the two controlled dimensions used in a given study as independent variables. Additionally, three-way ANOVA is used to investigate whether the controlled response dimensions and the question or user's familiarity with the topic have an effect on users' evaluation of the responses (measured with user-judged response dimensions). 
The results of our user studies are reported in~Section~\ref{sec:results}.
We analyze the crowdsourced data with the Python \texttt{statsmodels} library.\footnote{\url{https://www.statsmodels.org/}} We use significance level $\alpha=0.05$ to report statistical significance.
Whenever applicable, the $\omega^2$ unbiased effect size of a given factor is calculated to quantify the magnitude of the variance observed in the model. It is classified based on the scales used by~\citet{Culpepper:2022:ACM} ($\omega^2 \geq 0.14$: large effect size; $0.06$--$0.14$: medium; $0.01$--$0.06$: small; $\leq 0$: no effect).
\section{User Study Execution}
\label{sec:user_study_execution}

We used the Amazon Mechanical Turk (AMT)
 crowdsourcing platform to collect responses from online workers. 
The studies were run between 15 September 2023 and 4 October 2023.

\subsection{Data}
\label{subsec:data}
A critical element of the study is selecting query-response pairs that best represent the particular challenges.
We manually craft responses for twenty search queries from TREC CAsT'20~\citep{Dalton:2020:TREC} and '22~\citep{Owoicho:2022:TRECb},
\footnote{The TREC CAsT'19 dataset is less complex compared to the 2020 and 2022 editions, while the CAsT’21 dataset assesses relevance at the document level instead of passages.}
simulating everyday system interactions under various experimental conditions.
The responses are curated by the authors of the paper to ensure accordance with defined response dimensions and high data quality.

\subsubsection{\textbf{Queries}}
\label{subsubsec:queries}

For each user study, we select ten queries from the topics released in CAsT 2020 and 2022 that are susceptible to one of the identified problems (i.e., query answerability and response incompleteness) as detailed below.  

\myparagraph{Answerability Study}

To identify queries with unanswerability issues (i.e., queries for which answers have not been found), we use the information nugget (i.e., a piece of valuable information) annotations from the CAsT-snippets dataset~\citep{Lajewska:2023:CIKM} to indicate whether the answer or part of it has been found in the top retrieved passages. We aim to select queries not widely covered in the TREC CAsT passage collections and for which retrieving the answer was challenging. Based on the annotations provided in the CAsT-snippets dataset, we select queries that contain annotated snippets in some but not all of the top-5 passages (based on their ground truth relevance scores in the TREC CAsT datasets).
This way, we ensure that the query faces unanswerability problems, but some passages contain information that can be used to generate factually correct responses.
\footnote{Note that answerability can be determined w.r.t. a document (e.g., SQuAD 2.0~\citep{Rajpurkar:2018:ACL}), corpus (e.g., TREC CAsT~\citep{Dalton:2019:TREC}), knowledge base~\citep{PathiyanCherumanal:2024:CHIIR}, or  external expert knowledge. In this paper, we consider answerability w.r.t. a particular set of retrieved passages.}
After selecting potential candidates, we randomly select only one query per topic to maintain the study's topical diversity. The queries used in the \studyone{} are presented in Table~\ref{tab:queries_user_study_1}.

\begin{table}[tp]
    \centering
    \footnotesize
    \captionsetup{skip=5pt} 
    \caption{Queries from the TREC CAsT'20 and '22 datasets used in the \studyone{}.}
    \label{tab:queries_user_study_1}
    \begin{tabular}{@{~}l@{~~}l@{~}p{6.5cm}}
    \toprule
        \textbf{ID} & \textbf{TREC ID} & \textbf{Query} \\ \midrule
        1 & 146\_1-9 & What's the best bike seat \\
        2 & 135\_2-3 & How often should I run to lose weight? \\
        3 & 139\_2-15 & What are the other natural wonders of the world besides the Great Barrier Reef? \\
        4 & 142\_7-1 & I like hiking and Malbec wine. You mentioned some high peaks. How can I hike some high mountains and visit some wineries famous for Malbec? \\
        5 & 144\_2-11 & Tell me about the different types of rocket engines. \\
        6 & 147\_2-3 & Interesting. What was the basis of the backlash Marvel Studios faced for the Vice President’s suggestion that diversity was causing sales to slide? \\
        7 & 149\_3-1 & How do you get impartial results from search engines? \\
        8 & 82\_6 & What is the role of Co-Extra in GMO food traceability in the EU? \\
        9 & 85\_4 & What licenses and permits are needed for a food truck? \\
        10 & 90\_5 & Why did the Airbus A380 stop being produced? \\ \bottomrule
    \end{tabular}
    \vspace{-5pt}
\end{table}

\myparagraph{Viewpoints Study}

Open-ended queries about complex or contentious topics with multiple facets and/or viewpoints are specifically prone to incomplete responses \citep{Draws:2021:SIGIR}. To identify such queries in TREC CAsT collections, we: (1) manually select a subset of potential candidates and (2) ask crowdworkers to prioritize the selected queries in terms of their controversy and broadness. In step (1), we identify queries related to politics, society, environment, science, education, and technology. Queries strongly dependent on the conversational context or requiring background knowledge are not considered. In step (2), we run a small crowdsourcing task where workers are presented with a question and asked to assess its controversy and broadness on an ordinal scale of 1--5. Based on the collected judgments, we select the top 12 queries for which we generate different variants of the responses. 
At this stage, we select two additional queries to run an additional validation step (see~Section~\ref{subsubsec:responses} for more details about the process). The final ten queries used in the \studytwo{} are presented in Table~\ref{tab:queries_user_study_2}.

\begin{table}[tp]
    \centering
    \footnotesize
    \captionsetup{skip=5pt} 
    \caption{Queries from the TREC CAsT'20 and '22 datasets used in the \studytwo{}.    \label{tab:queries_user_study_2}}
    \begin{tabular}{@{~}l@{~~}l@{~}p{6.8cm}}
    \toprule
        \textbf{ID} & \textbf{TREC ID} & \textbf{Query} \\ 
        \midrule
        1 & 137\_1-5 & What do other philosophers think about Bostrom's `simulation argument'? \\   
        2 & 105\_6 & What was the US reaction to the Black Lives Matter movement? \\ 
        3 & 102\_8 & Can social security be fixed? \\
        4 & 149\_2-5 & Are algorithms really biased against people of colour \\
        5 & 136\_1-13 & What effects did the Watergate scandal have on President Nixon's legacy? \\
        6 & 138\_1-9 & Do you think social media might play a role in my son's low self-esteem? \\
        7 & 91\_7 & What do users of social networks get in return for by giving up their privacy? \\
        8 & 147\_2-1 & What is Marvel Studios’ approach to diversity for people of color? \\
        9 & 82\_2 & What are the pros and cons of GMO food labeling? \\
        10 & 132\_2-1 & That’s interesting. Tell me more about how climate change affects developing countries. \\ \bottomrule
    \end{tabular}
    \vspace{-5pt}
\end{table}

\subsubsection{\textbf{Responses}}
\label{subsubsec:responses}

The responses were manually created by the authors of this paper and are based on the five most relevant passages in the TREC CAsT datasets. 
The selected passages were first summarised using GPT-3.5, then manually reviewed and embellished to add or remove information, verify the correctness, introduce factual errors, or balance the content depending on the experimental condition.
We identify two main dimensions for generating system responses in each user study, acknowledging that these dimensions are not exhaustive.
Nevertheless, our hypothesis posits that varying the responses along these dimensions will give us the means to answer our research questions effectively.

\myparagraph{Answerability Study}

\begin{table*}[tp]
    \centering
    \footnotesize
    \captionsetup{skip=5pt} 
    \caption{
    Schema for experimental conditions ($EC^A_1$--$EC^A_4$) in the \studyone{}. The last two columns contain different variants of CIS system response along with the source for Query 4 (cf. Table~\ref{tab:queries_user_study_1}).    }
    \label{tab:answers_unanswerability}
    \adjustbox{max width=0.95\textwidth}{
    \begin{tabular}{cp{3cm}ccp{7cm}p{6.4cm}}
    \toprule
          & \textbf{Experimental} & \multicolumn{2}{c}{\textbf{Response Dimension}} & \multirow{2}{*}{\textbf{CIS System Response}} & \multirow{2}{*}{\textbf{Source}}  \\
         \cmidrule{3-4}
          & \textbf{Condition} & \textbf{Factual Corr.} & \textbf{Source}     \\
        \midrule
        \multirow{3}{*}{\(EC^A_1\)} & \multirow{3}{3cm}{Factually correct + valid source} & \multirow{3}{*}{\checkmark} & \multirow{3}{*}{\checkmark} & \emph{You can combine your love for hiking and Malbec wine by visiting Mendoza, Argentina. This picturesque city is nestled in the Andes and is renowned for its vineyards...} & \url{https://wanderingtrader.com/argentina/top-5-argentina-tourist-attractions/}   \\
        \midrule
        \(EC^A_2\) & {Factually correct + no source} & \checkmark & \cross & Same as above & -- \\
        \midrule
        \multirow{4}{*}{\(EC^A_3\)} & \multirow{4}{3cm}{Factually incorrect + invalid source} & \multirow{4}{*}{\cross} & \multirow{4}{*}{\checkmark (invalid)} & \emph{To combine hiking and Malbec wine, plan a trip to the Penedès region in Catalonia. You can explore the Montserrat mountain range, which offers fantastic hiking opportunities, and then visit renowned wineries in the Penedès, known for its exceptional Malbec wine production\ldots} & \url{https://www.winetourism.com/wine-tasting-tours-in-penedes/} \newline (The link is valid but the article is a website with Wine Tasting \& Tours in Penedès, Spain where Malbec wine is not produced.) \\
        \midrule
        \(EC^A_4\) & Factually incorrect + no source & \cross & \cross & Same as above & -- \\
        \bottomrule
    \end{tabular}
    }
    \vspace{-5pt}
\end{table*}

Failure to find the exact answer to the query in CIS can lead to factual errors and hallucinations (i.e., the introduction of facts that are not true). This is a common problem especially when the response is generated as a summary of partially relevant passages using large language models~\citep{Tang:2023:ACL}.
Therefore, we are mostly interested in the following two response dimensions:
\begin{enumerate}
    \item factual correctness of the included information, and
    \item the presence and validity of the source of the information.
\end{enumerate}
The accurate response contains factually correct information along with the source (\(EC^A_1\)).
Whereas, the flawed response fails to provide a source (\(EC^A_2\)), contains factually incorrect/unsupported information with an invalid source (\(EC^A_3\)), or without a source (\(EC^A_4\)); see~Table~\ref{tab:answers_unanswerability}. The flawed response may contain various factual inconsistencies, such as negation and number, entity, or antonym swaps~\citep{Kryscinski:2020:EMNLP}, as well as fully hallucinated content not supported by any source information~\citep{Ji:2023:ACMa, Liu:2023:EMNLP}. An invalid source indicates a mismatch between the source's name and content, a topically relevant source that does not support the specific facts in the response, or a source with a corrupted link. Following the setup proposed for evaluating the usefulness of supporting documents in the WikiHowQA benchmark~\citep{Bolotova-Baranova:2023:ACL}, we allow workers to freely examine the sources linked in the responses to evaluate their correctness and relevance. 

\myparagraph{Viewpoints Study}

\begin{table*}[tp]
    \centering
    \footnotesize
    \captionsetup{skip=5pt} 
    \caption{Schema for experimental conditions ($EC^V_1$--$EC^V_3$) in the \studytwo{}. 
    The last column contains different variants of CIS system response for Query 5 (cf.~Table~\ref{tab:queries_user_study_2}).}
    \adjustbox{max width=0.95\textwidth}{
    \begin{tabular}{lp{2.1cm}ccp{16.5cm}}
    \toprule
          & \textbf{Experimental} & \multicolumn{2}{l}{\textbf{Response Dimension}} & \multirow{2}{*}{\textbf{CIS System Response}} \\
         \cmidrule{3-4}
          & \textbf{Condition} & \textbf{Diversity} & \textbf{Balance} & \\
        \midrule
        \multirow{3}{*}{\(EC^V_1\)} & \multirow{3}{2.1cm}{Multiple viewpoints covered to the same extent} & \multirow{3}{*}{\checkmark} & \multirow{3}{*}{\checkmark} & \emph{The Watergate scandal initially had a severely damaging effect on President Nixon's legacy, tarnishing his image as a corrupt politician and overshadowing his accomplishments. However, in later years, Nixon's reputation underwent a degree of rehabilitation, with recognition of his roles in civil rights and diplomacy. He transitioned from being primarily associated with scandal to being acknowledged as a respected elder statesman, although his legacy remains marked by the shadow of Watergate.} \\
        \midrule
        \multirow{4}{*}{\(EC^V_2\)} & \multirow{4}{2.1cm}{Multiple viewpoints mentioned, one covered in detail} & \multirow{4}{*}{\checkmark} & \multirow{4}{*}{\cross} & \emph{The Watergate scandal had a profoundly negative impact on President Nixon's legacy, overshadowing many of his domestic achievements. It tarnished his reputation as a corrupt politician, making him a symbol of political scandal and misconduct in both American politics and popular culture. While he did enact significant legislation like creating the Environmental Protection Agency, his presidency is primarily remembered for the scandal, leading to his resignation and becoming synonymous with political corruption and disgrace. } \\
        \midrule
        \multirow{3}{*}{\(EC^V_3\)} & \multirow{3}{2.1cm}{Single viewpoint mentioned and covered} & \multirow{3}{*}{\cross} & \multirow{3}{*}{\cross} & \emph{The Watergate scandal had a profoundly negative impact on President Nixon's legacy. It tarnished his reputation as a corrupt politician, making him a symbol of political scandal and misconduct in both American politics and popular culture. Nixon's resignation and the scandal's fallout reinforced public skepticism and criticism of the presidency, leaving a lasting impression as one of the most Shakespearean and disgraceful episodes in presidential history.} \\
        \bottomrule
    \end{tabular}
    }
    \label{tab:answers_incompletenes}
\end{table*}

Research on debated topics typically represents viewpoints in a binary fashion (in favor/against). However, viewpoints are additionally characterized by stance, i.e., the degree of strength (e.g., slight support vs. strong favor) and the logic of evaluation (underlying reason or perspective behind the stance)~\citep{Draws:2022:CHIIRa}. 
Our user study does not address the stance or evaluation logic and focuses on a widely understood diversity of viewpoints and facets. Crowd workers are asked to judge whether the expressed viewpoints or described topic facets are diverse enough or not. While investigating queries that are likely to result in incomplete responses, we are interested in the following two dimensions:
\begin{enumerate}
    \item response diversity in terms of different viewpoints and/or facets mentioned, and
    \item balance in the amount of information provided for each viewpoint and/or facet.
\end{enumerate}
\vspace{-2pt}
The accurate response equally covers various points of view and/or facets of the topic to the same extent (\(EC^V_1\)). 
The flawed response mentions several viewpoints and/or topic facets but elaborates only on one of them (\(EC^V_2\)) or mentions only one (\(EC^V_3\)); see~Table~\ref{tab:answers_incompletenes}.
\footnote{Note that a text discussing a single viewpoint or facet cannot be unbalanced; therefore, an experimental condition with a lack of diversity and balance is not applicable.}

We introduce an additional step for the \studytwo{} to validate our proposed response dimensions: diversity, and balance. 
This step, addressing the subjectivity of controversy and topic broadness, aids in filtering out non-representative query-response pairs.
We create small surveys where expert annotators are presented with three topics and lists of recommended resources used to generate the responses. Expert annotators are asked to explore the provided resources to become familiar with the given topic. 
Then, they are presented with different response variants and asked to judge the diversity and balance of each of the provided query-response pairs. 
For each of the twelve queries, we collect ratings between 1--5 for diversity and balance from three different expert annotators.
We employ Ph.D. students for their academic skills in exploring new domains, assuming their ratings reflect users highly familiar with the topics (i.e., experts).
We exclude the query for which the response variant corresponding to \(EC^V_1\) (multiple viewpoints covered to the same extent) is judged as not diverse enough and the query for which the response variant corresponding to \(EC^V_3\) (single viewpoint mentioned and covered) is judged as too balanced.

\subsection{Workers}
\label{subsec:Workers}
Crowd workers with an approval rate greater than 97\%, more than 5,000 approved HITs, and located in the US were qualified to participate in the studies. 
Workers were paid \$3 USD for successful HIT completion. The reward was estimated based on the time needed by an expert to complete the task (the time was increased by 30\%) and the federal minimum wage in the US (\$7.25 USD per hour). 
Three different workers assessed each query-response pair to avoid repeated judgments that would reduce the reliability of the study~\citep{Steen:2021:ACL}. This user study setup gave us 12 (3 workers $\times$ 4 answer variants per query) different HITs for the \studyone{} and 9 (3 workers $\times$ 3 answer variants per query) for the \studytwo{}. This resulted in 36 annotators for \studyone{} and 27 annotators for \studytwo{}. The power analysis~\footnote{Calculated using the scripts  at~\url{https://waseda.app.box.com/v/SIGIR2016PACK}}, employing results of one-way ANOVA with the experimental condition as an independent variable and the user-reported values for the main response dimension (factual correctness for the \studyone{} and diversity for the \studytwo{}) as a dependent variable, was conducted using data collected in the first run. 
The results of the power analysis indicated that \studytwo{} had a strong ``true'' effect when it existed. In contrast, the low power of \studyone{} suggested a low statistical sensitivity. 
-- aligning with our intuition and previous research, which suggests that users are unlikely to detect hallucinations~\citep{Lajewska:2024:SIGIR}. To increase the power of \studyone{}, we collected more data from five additional workers per HIT in the second run with the same worker requirements and rewards (see~Table~\ref{tab:user_studies_numbers} for descriptive statistics).
Ten submissions out of 133 released HITs were discarded due to failed attentiveness checks.

The \studyone{} involved 96 workers: 44 male and 52 female (no workers reported ``other'' or ``prefer not to say'').
Thirty-four workers self-reported to be in the 18--30 age group, 35 in the 31--45 group, 19 in the 46--60, and seven in the 60+ group. One participant did not report on age. 
Regarding education, two workers held a Ph.D. or higher, 15 had a master's degree, 59 had a bachelor's degree, and 19 had high school. One participant did not report on education.
The \studytwo{} involved 27 workers: 15 male and 12 female (with none selecting "Other" or "Prefer not to say").
Three workers self-reported to be in the 18--30 age group, 12 in the 31--45 group, 10 in the 46--60, and two in the 60+ group.
Two workers had a master's degree, 16 had a bachelor's degree, and 8 had a high school education. One participant did not report on education. 

\begin{table}[tp]
    \centering
    \footnotesize
    \captionsetup{skip=5pt} 
    \caption{User studies setup in numbers. Numbers in the parentheses refer to the second data collection run.}
    \label{tab:user_studies_numbers}
    \begin{tabular}{@{~}l@{}cc}
       \toprule
         & \textbf{Answerability} & \textbf{Viewpoints} \\
        \midrule
        \#queries per user study & 10 & 10 \\
        \#experimental cond. (\#resp. per query) & 4 & 3 \\
        \#crowd workers per HIT & 3 (+5) & 3 \\
        \#different HITs & 12 & 9 \\
        \#crowd workers per query-response & 9 (+15) & 9 \\
        \#query-response pairs annotations & 360 (+600) & 270 \\
        \bottomrule
    \end{tabular}
\end{table}
\section{Results and Discussion}
\label{sec:results}

The analysis of data obtained from the crowdsourcing experiments is performed using the methods described in~Section~\ref{subsec:data_analysis_methods}. 

\begin{table}[tp]
\captionsetup{skip=5pt} 
\caption{Results of two-way ANOVA. Statistically significant effects are in bold. Effect size: L=Large, M=Medium, S=Small.}
    \label{tab:two-way-anova}
    \centering
    \footnotesize
    \newlength{\depvarlength}
    \setlength{\depvarlength}{1.6cm}
    \sisetup{
    detect-weight=true
    }
    \begin{tabular}{p{\depvarlength}cS[table-format=1.3,table-auto-round]c}
    \toprule
           {\begin{tabular}   {@{}l@{}}\textbf{Dependent Variable}\\\textbf{(User-Judged)}
        \end{tabular}} & {\begin{tabular}   {@{}l@{}}\textbf{Independent Variable(s)}\\\textbf{(Controlled)} \end{tabular}} & \textbf{$p$-value} & {\begin{tabular}   {@{}c@{}}\textbf{Effect}\\\textbf{Size}
        \end{tabular}} \\
        \midrule
        \multicolumn{2}{l}{\titlecap{\studyone}} \\
        \midrule
        \multirow{3}{*}{Factual Correctness} & \textbf{Contr. Fact. Corr.} & \textbf{0.014} & \textbf{-} \\
         & Contr. Source &  0.664 & - \\
         & Contr. Fact. Corr. * Contr. Source &  0.267 & - \\
        \midrule
        \multirow{3}{*}{Conf. in Answer Acc.} & Contr. Fact. Corr. & 0.244 & - \\
         & Contr. Source &  0.763 & - \\
         & Contr. Fact. Corr. * Contr. Source &  0.575 & - \\
        \midrule
        \multirow{3}{*}{Overal Satisfaction} & Contr. Fact. Corr. & 0.306 & - \\
         & Contr. Source &  0.394 & - \\
         & Contr. Fact. Corr. * Contr. Source &  0.267 & - \\
        \midrule
        \multicolumn{2}{l}{\titlecap{\studytwo}} \\
        \midrule
        \multirow{3}{*}{Diversity} & \textbf{Contr. Diversity} & \textbf{0.0} & \textbf{M} \\
         & Contr. Balance & 1.0 & - \\
         & \textbf{Contr. Diversity * Contr. Balance} & \textbf{0.0} & \textbf{M} \\
        \midrule
        \multirow{3}{*}{Transparency} & \textbf{Contr. Diversity} & \textbf{0.0} & \textbf{M} \\
         & Contr. Balance & 1.0 & - \\
         & \textbf{Contr. Diversity * Contr. Balance} & \textbf{0.0} & \textbf{M} \\
        \midrule
        \multirow{3}{*}{Balance} & \textbf{Contr. Diversity} & \textbf{0.0} & \textbf{S} \\
         & Contr. Balance & 1.0 & - \\
         & \textbf{Contr. Diversity * Contr. Balance} & \textbf{0.0} & \textbf{S} \\
        \midrule
        \multirow{3}{*}{Overall Satisfaction} & \textbf{Contr. Diversity} & \textbf{0.0} & \textbf{S} \\
         & Contr. Balance & 1.0 & - \\
         & \textbf{Contr. Diversity * Contr. Balance} & \textbf{0.0} & \textbf{M} \\
       \bottomrule
    \end{tabular}

\end{table}

\begin{table}[tp]
\captionsetup{skip=5pt} 
\caption{Results of one-way ANOVA. Statistically significant effects are in bold. Effect size: L=Large, M=Medium, S=Small.}
    \label{tab:one-way-anova}
    \centering
    \footnotesize
    \setlength{\depvarlength}{2.5cm}
    \sisetup{
    detect-weight=true
    }
    \begin{tabular}{p{\depvarlength}cS[table-format=1.3,table-auto-round]c}
    \toprule
           {\begin{tabular}   {@{}l@{}}\textbf{Dependent}\\\textbf{Variable}
        \end{tabular}} & {\begin{tabular}   {@{}l@{}}\textbf{Independent}\\\textbf{Variable(s)} \end{tabular}} & \textbf{$p$-value} & {\begin{tabular}   {@{}c@{}}\textbf{Effect}\\\textbf{Size}
        \end{tabular}} \\
        \midrule
        \multicolumn{2}{l}{\titlecap{\studyone}} \\
        \midrule
        \textbf{Familiarity} & \multirow{4}{*}{Query} & \textbf{0.0} & \textbf{M} \\
        \textbf{Factual Corr.} &  & \textbf{0.0} & \textbf{S} \\
        \textbf{Conf. in Answer Acc.} &  & \textbf{0.019} & \textbf{S} \\
        \textbf{Overall Satisfaction} &  & \textbf{0.0} & \textbf{S} \\
        \midrule
        \textbf{Factual Correctness} & \multirow{3}{*}{Familiarity} & \textbf{0.005} & \textbf{S} \\
        \textbf{Conf. in Answer Acc.} &  & \textbf{0.0} & \textbf{S} \\
        \textbf{Overall Satisfaction} &  & \textbf{0.0} & \textbf{M} \\
        \midrule
        \multicolumn{2}{l}{\titlecap{\studytwo}} \\
        \midrule
        \textbf{Familiarity} & \multirow{5}{*}{Query} & \textbf{0.0} & \textbf{L} \\
        Diversity &  & 0.338 & - \\
        Transparency &  & 0.458 & - \\
        \textbf{Balance} &  & \textbf{0.027} & \textbf{S} \\
        \textbf{Overall Satisfaction} &  & \textbf{0.005} & \textbf{S} \\
        \midrule
        Diversity & \multirow{4}{*}{Familiarity} & 0.375 & - \\
        transparency &  & 0.478 & - \\
        Balance &  & 0.639 & - \\
        Overall Satisfaction &  & 0.378 & - \\
       \bottomrule
    \end{tabular}
    \vspace{-5pt}
\end{table}

\begin{table}[tp]
\captionsetup{skip=5pt}
\caption{Results of three-way ANOVA. Stat. significant effects are in bold. Effect size: L=Large, M=Medium, S=Small.}
    \label{tab:three-way-anova}
    \centering
    \footnotesize
    \setlength{\depvarlength}{1.6cm}
    \sisetup{
    detect-weight=true
    }
    \adjustbox{max width=0.46\textwidth}{
    \begin{tabular}{p{\depvarlength}cS[table-format=1.3,table-auto-round]c}
    \toprule
           {\begin{tabular}   {@{}l@{}}\textbf{Dependent Variable}\\\textbf{(User-Judged)}
        \end{tabular}} & {\begin{tabular}   {@{}l@{}}\textbf{Independent Variable(s)}\\\textbf{(Controlled)} \end{tabular}} & \textbf{$p$-value} & {\begin{tabular}   {@{}c@{}}\textbf{Effect}\\\textbf{Size}
        \end{tabular}} \\
        \midrule
        \multicolumn{2}{l}{\titlecap{\studyone}} \\
        \midrule
        \multirow{4}{*}{Factual Correctness} & \textbf{Query} & \textbf{0.0} & \textbf{S} \\
         & \textbf{Contr. Fact. Corr. * Query} & \textbf{0.002} & \textbf{S} \\
         & \textbf{Contr. Source * Query} & \textbf{0.048} & \textbf{-} \\
         & Contr. Fact. Corr. * Contr. Source * Query & 0.439 & - \\
        \midrule
        \multirow{4}{*}{Conf. in Answer Acc.} & \textbf{Query} & \textbf{0.015} & \textbf{S} \\
         & \textbf{Contr. Fact. Corr. * Query} & \textbf{0.0} & \textbf{S} \\
         & Contr. Source * Query & 0.118 & - \\
         & Contr. Fact. Corr. * Contr. Source * Query & 0.341 & - \\
        \midrule
        \multirow{4}{*}{Overall Satisfaction} & \textbf{Query} & \textbf{0.0} & \textbf{S} \\
         & \textbf{Contr. Fact. Corr. * Query} & \textbf{0.0} & \textbf{S} \\
         & Contr. Source * Query & 0.339 & - \\
         & Contr. Fact. Corr. * Contr. Source * Query & 0.598 & - \\
        \midrule
        \multicolumn{2}{l}{\titlecap{\studytwo}} \\
        \midrule
        \multirow{4}{*}{Diversity} & Query & 0.147 & S \\
         & Contr. Diversity * Query & 0.101 & S \\
         & Contr. Balance * Query & 1.0 & - \\
         & \textbf{Contr. Diversity * Contr. Balance * Query} & \textbf{0.016} & \textbf{S} \\
        \midrule
        \multirow{4}{*}{Transparency} & Query & 0.35 & - \\
         & Contr. Diversity * Query & 0.582 & - \\
         & Contr. Balance * Query & 1.0 & - \\
         & Contr. Diversity * Contr. Balance * Query & 0.689 & - \\
        \midrule
        \multirow{4}{*}{Balance} & \textbf{Query} & \textbf{0.012} & \textbf{S} \\
         & Contr. Diversity * Query & 0.559 & - \\
         & Contr. Balance * Query & 1.0 & - \\
         & Contr. Diversity * Contr. Balance * Query & 0.316 & - \\
        \midrule
        \multirow{4}{*}{Overall Satisfaction} & \textbf{Query} & \textbf{0.001} & \textbf{M} \\
         & Contr. Diversity * Query & 0.599 & - \\
         & Contr. Balance * Query & 1.0 & - \\
         & \textbf{Contr. Diversity * Contr. Balance * Query} & \textbf{0.034} & \textbf{S} \\
       \bottomrule
    \end{tabular}
     }
\end{table}

\subsection{Users' Ability to Recognize Problems} 
\label{sec:results_rq1}

Table~\ref{tab:two-way-anova} shows the results of the two-way ANOVA performed to answer RQ1 (\emph{Can users effectively recognize problems related to query answerability and response incompleteness in system responses?}). Controlled response dimensions are treated as independent variables, and a given response dimension (i.e., self-reported worker ratings) as a dependent variable. Statistically significant results indicate an effect of the experimental condition on a given response dimension.

\myparagraph{Effect of controlled response dimension manipulation on response user ratings}

We do not observe any statistically significant effect of manipulating the controlled response dimensions on user ratings in the \studyone{} (upper part of Table~\ref{tab:two-way-anova}), suggesting that users cannot recognize pitfalls in the responses or do not associate them with any of the response dimensions. 
On the other hand, results for the \studytwo{} (lower part of Table~\ref{tab:two-way-anova}) show small or medium effect on self-reported worker ratings meaning that users can correctly identify the problems related to viewpoint diversity and balance.

\myparagraph{Effect of the interaction between query and controlled response dimensions on user ratings}

The three-away ANOVA results in Table~\ref{tab:three-way-anova} show that the query and interaction between the query and the controlled response dimensions (especially factual correctness) significantly affect all response dimensions in the \studyone{}, which aligns with findings from other information retrieval experiments, highlighting the topic-dependent nature of user judgments~\cite{Culpepper:2022:ACM,Alaofi:2022:SIGIR}.
It indicates that the perceived factual correctness may vary based on the query, despite the consistent experimental condition. 
In the \studytwo{}, only the diversity and overall satisfaction with the response are affected by the interaction between the query and controlled response dimensions, suggesting that the \studytwo{} is more robust w.r.t. topic/query variability.

\myparagraph{Effect of the interaction between user background knowledge and experimental condition}

The topic familiarity reported by workers is a proxy for user background knowledge. Even though we anticipated that the topic familiarity would influence the ratings reported by the workers for different response dimensions, we did not observe a statistically significant association of the interaction between the familiarity and experimental condition on any of the response dimensions. This holds for both user studies.\footnote{Detailed results are presented in Appendix~\ref{app:anova-familiarity}.}

\subsection{User Experience}
\label{sec:results_rq2}

This section discusses the results to answer RQ2 (\emph{How do factually incorrect, inaccurate, incomplete, and/or biased responses impact the user experience?)}.

\begin{table}[tp]
    \centering
    \captionsetup{skip=5pt}
    \caption{
    Pearson correlation between user-reported response dimensions and their overall satisfaction with system's response.}
    \label{tab:pearson}
    \footnotesize
    \begin{tabular}{lS[detect-weight]}
    \toprule
         \textbf{Response Dimension} & \textbf{Correlation Coefficient} \\
        \midrule
        \titlecap{\studyone} \\
        \midrule
        Factual Correctness & 0.634 \\
        Conf. in Answer Acc. & 0.660 \\
        \midrule
        \titlecap{\studytwo} \\
        \midrule
        Diversity & 0.720\\
        Transparency & 0.727 \\
        Balance & 0.785 \\
        \bottomrule
    \end{tabular}
\vspace{-10pt}
\end{table}

\myparagraph{Correlation between user-reported response dimensions and the overall satisfaction}

Table~\ref{tab:pearson} shows the Pearson correlation coefficient $r$ calculated for overall satisfaction---a proxy for user experience---, and user-reported response dimensions. For both user studies, we observe a moderately strong correlation ($0.6<r<0.8)$ between user satisfaction and other user-judged dimensions.  This suggests that satisfaction is a fairly good indicator of overall user experience. Correlations for the \studyone{} are lower than for the \studytwo{}. As we discussed in Section~\ref{sec:results_rq1}, we do not observe a statistically significant effect of the controlled response dimension on user ratings for the \studyone{}. This implies that users find these response dimensions important and associate them with their satisfaction, but they are not able to identify them correctly in system responses. On the other hand, 
results for the \studytwo{} suggest that users can correctly identify these dimensions and use them as indicators for their satisfaction.

\myparagraph{Effect of query and response quality on overall satisfaction}

In both studies, the query significantly affects overall satisfaction (see~Table~\ref{tab:one-way-anova}). We do not observe a statistically significant association between controlled response dimensions and overall satisfaction in the \studyone{}, which suggests that response quality does not influence worker's perception of satisfaction (see~Table~\ref{tab:two-way-anova}).
The opposite observation is made in the \studytwo{}, implying that workers can spot response inaccuracies.
The three-way ANOVA (Table~\ref{tab:three-way-anova}) shows that a small- or medium-size effect of the query leads to a statistically significant effect of the interaction between query and response variant on the overall satisfaction for both studies. This indicates that, in terms of user satisfaction, both studies are sensitive to topic variability that may impact the results.
For future work, using a larger number of queries, especially for \studyone{}, may increase the sensitivity of the experiment.

\subsection{Further Analysis}
\label{sec:results_further_analysis}

\myparagraph{Rating distributions for response dimensions}

In the \studyone{}, the ratings for user-judged response dimensions, topic familiarity, and overall satisfaction per query are concentrated around higher values (3 and 4) for all response dimensions apart from familiarity.
\footnote{Data distribution can be found in Appendix~\ref{app:data_distribution}.} 
It means that workers are not very critical in evaluating these dimensions or cannot identify the pitfalls related to them. Workers report that they are rather unfamiliar with most of the query topics. In the \studytwo{}, the ratings for familiarity are more spread. A wide range of diversity ratings is observed per query, unlike for other response dimensions. Even though the ratings are more spread than for the \studyone{}, most of the ratings concentrate around a higher value (i.e., 3).

\myparagraph{Effect of background knowledge on the response dimensions}

According to the results of one-way ANOVA with familiarity used as an independent variable (see Table~\ref{tab:one-way-anova}), we obtain different results for the two studies. In the \studyone{}, 
the worker's background knowledge impacts how accurate or satisfying they find the response.
Whereas, in the \studytwo{}, none of the response dimensions is significantly affected by users' topic familiarity. 

\myparagraph{Effect of the query on the response dimensions}

In both user studies the topic familiarity and overall user satisfaction are significantly affected by the query (see Table~\ref{tab:one-way-anova}).
It means that user background knowledge and response satisfaction depend on the query, not necessarily on the response. It confirms that, to get meaningful results, one must include many different study topics, which is indeed what we tried to ensure with our query selection processes. Statistically significant differences in response dimensions between queries are observed for all dimensions in the \studyone{}, while only for balance in the \studytwo{}. This suggests that the former studies' setup is more query-dependent than the latter. The results are more generalizable in the \studytwo{}, even after collecting additional data according to the power analysis results for the \studyone{}. The high effect of the query on all the response dimensions in the \studyone{} also justifies the significant effects of the interactions between the query and the controlled response dimensions observed in the three-way ANOVA.
\section{Discussion}
\label{sec:discussion}

Users generally find it easier to perceive viewpoints than to assess factual correctness. 
In the \studyone{}, crowd workers demonstrate a limited ability to detect pitfalls in responses compared to the \studytwo{}, highlighting the challenge of identifying factual errors without topic-specific knowledge.
In terms of user satisfaction, in the \studyone{} it strongly correlates with confidence in answer accuracy, highlighting the importance of valid sources. In the \studytwo{}, satisfaction is tied to perceived balance, with users preferring unbiased responses that equally cover all viewpoints.
Satisfaction scores reported by users do not always align with their comments---additional aspects revealed in free-text user comments refer to source credibility, as well as the completeness, usefulness, and subjectivity of the provided information---, indicating a potential discrepancy between reported and actual satisfaction levels.\footnote{Additional qualitative analysis of the impact of response inaccuracies and biases on user experience based on free-text comments is presented in Appendix~\ref{app:qualitative_analysis}.}
Users may also associate their satisfaction with response fluency, that can be easily ensured by existing generative search engines. However, it does not guarantee the accuracy or proper citation of all statements~\citep{Liu:2023:EMNLP}.

The conclusions drawn from these studies inform the design of future response generation methods and highlight important challenges that still need to be addressed. Simply relying on the relevance of the top retrieved passages does not guarantee the generation of a satisfying response. Future response generation approaches must ensure the completeness, diversity, balance, objectivity, and factual correctness of responses, along with proper attribution to credible sources.
Additionally, the response should inform users of potential inaccuracies and help them assess the presented information objectively, by providing sources or system capability details. Including these explanations ensures transparent and effective interactions with the system~\citep{Lajewska:2024:SIGIR}.
Another open challenge is the evaluation of the generated responses. To the best of our knowledge, there are no CIS datasets with ground truth judgments for the identified response dimensions. 
Our study designs and experimental protocol can serve as a blueprint for human evaluation of responses across multiple dimensions, supporting data collection for a broader range of experimental conditions, more complex multi-turn settings, and additional queries/topics.

\myparagraph{Limitations}

Due to the complexity of the user studies and the costs involved, some simplifications were made, such as focusing on single-turn interactions and limited number of queries. As a result, these experiments do not fully reflect the dynamic nature of real-world CIS dialogues, where user needs and context change over multiple turns.
Future work will explore more topics, particularly for the \studyone{}, to enhance result sensitivity, and use other scales to capture overall satisfaction (e.g., magnitude estimation~\citep{Turpin:2015:SIGIR}). Another limitation is relying on Amazon MTurk crowd workers, who may not fully represent the diversity of CIS system users. These studies do not fully control participants own biases, which is left for future investigation.
Lastly, the findings of this work are limited to the properties of the test collection used in our experiments. Future experiments should also explore answerability on broader levels---such as ranking, corpus, and expert knowledge---while considering the system's transparency when no answer is found, as well as a wider spectrum of topics, viewpoints, and responses.
Despite these limitations, the experiments serve as a first step toward understanding challenges in CIS response generation and highlight key open questions for further research.
\section{Conclusions}
\label{sec:conclusions}

Response generation poses various challenges in CIS systems. To study this, we proposed two crowdsourcing-based study designs to investigate unanswerable questions and incomplete responses from a user perspective in the scenario inspired by the TREC CAsT benchmark. We explored users' ability to recognize factual inaccuracies, pitfalls, and biases in terms of viewpoint diversity by controlling experimental conditions in manually crafted responses simulating CIS system interactions.
Our findings provide evidence that: (i) CIS system responses cannot be limited to a simple synthesis of the retrieved information; and (ii) source attribution alone is insufficient to ensure effective interaction with the system.
We believe CIS responses should explicitly inform users about potential inaccuracies and provide aid to assess the presented information objectively (e.g., by including credible sources or information about system capabilities). 
The results presented in this paper can be regarded as guidelines for designing CIS solutions and conducting a more comprehensive analysis of the problems in the future.\footnote{Appendix~\ref{app:lessons} presents lessons learned and recommendations for conducting user studies that could be broadly useful for the community.}

\begin{acks}
    This research was supported by the Norwegian Research Center for AI Innovation, \grantsponsor{NorwAI}{NorwAI}{https://www.ntnu.edu/norwai} (Research Council of Norway, nr.~\grantnum{NorwAI}{309834}), and by the \grantsponsor{ARC}{Australian Research Council}{https://www.arc.gov.au/} (\grantnum{ARC}{DE200100064}, \grantnum{ARC}{CE200100005}).
\end{acks}
\newpage
\balance
\bibliographystyle{ACM-Reference-Format}
\bibliography{sigirap2024-resgen}

\newpage
\appendix
\section{Effect of User's Familiarity with the Topic}
\label{app:anova-familiarity}

Table~\ref{tab:three-way-anova-familiarity} presents the results of three-way ANOVA investigating the effect of the controlled response dimensions and user’s familiarity with the topic on users’ evaluation of the responses (measured with user-judged response dimensions).

\begin{table}[tp]
\captionsetup{skip=5pt}
\caption{Results of three-way ANOVA. Stat. significant effects are in bold. Effect size: L=Large, M=Medium, S=Small.}
    \label{tab:three-way-anova-familiarity}
    \centering
    \footnotesize
    \setlength{\depvarlength}{1.6cm}
    \sisetup{
    detect-weight=true
    }
    \adjustbox{max width=0.46\textwidth}{
    \begin{tabular}{p{\depvarlength}cS[table-format=1.3,table-auto-round]c}
    \toprule
           {\begin{tabular}   {@{}l@{}}\textbf{Dependent Variable}\\\textbf{(User-Judged)}
        \end{tabular}} & {\begin{tabular}   {@{}l@{}}\textbf{Independent Variable(s)}\\\textbf{(Controlled)} \end{tabular}} & \textbf{$p$-value} & {\begin{tabular}   {@{}c@{}}\textbf{Effect}\\\textbf{Size}
        \end{tabular}} \\
        \midrule
        \multicolumn{2}{l}{\titlecap{\studyone}} \\
        \midrule
        \multirow{4}{*}{Fact. Corr.} & \textbf{Familiarity} & \textbf{0.006} & \textbf{S} \\
         & Contr. Fact. Corr. * Familiarity & 0.962 & -- \\
         & Contr. Source * Familiarity & 0.275 & -- \\
         & Contr. Fact. Corr. * Contr. Source * Familiarity & 0.56 & -- \\
        \midrule
        \multirow{4}{*}{Conf. in Answer Acc.} & \textbf{Familiarity} & \textbf{0.0} & \textbf{S} \\
         & Contr. Fact. Corr. * Familiarity & 0.894 & -- \\
         & Contr. Source * Familiarity & 0.556 & -- \\
         & Contr. Fact. Corr. * Contr. Source * Familiarity & 0.348 & -- \\
        \midrule
        \multirow{4}{*}{Overall Satisfaction} & \textbf{Familiarity} & \textbf{0.0} & \textbf{M} \\
         & Contr. Fact. Corr. * Familiarity & 0.544 & -- \\
         & Contr. Source * Familiarity & 0.381 & -- \\
         & Contr. Fact. Corr. * Contr. Source * Familiarity & 0.777 & -- \\
        \midrule
        \multicolumn{2}{l}{\titlecap{\studytwo}} \\
        \midrule
        \multirow{4}{*}{Diversity} & Familiarity & 0.816 & -- \\
         & Contr. Diversity * Familiarity & 0.056 & S \\
         & Contr. Balance * Familiarity & 1.0 & -- \\
         & Contr. Diversity * Contr. Balance * Familiarity & 0.628 & -- \\
        \midrule
        \multirow{4}{*}{Transparency} & Familiarity & 0.788 & -- \\
         & Contr. Diversity * Familiarity & 0.257 & -- \\
         & Contr. Balance * Familiarity & 1.0 & -- \\
         & Contr. Diversity * Contr. Balance * Familiarity & 0.316 & -- \\
        \midrule
        \multirow{4}{*}{Balance} & Familiarity & 0.89 & -- \\
         & Contr. Diversity * Familiarity & 0.325 & -- \\
         & Contr. Balance * Familiarity & 1.0 & -- \\
         & Contr. Diversity * Contr. Balance * Familiarity & 0.242 & -- \\
        \midrule
        \multirow{4}{*}{Overall Satisfaction} & Familiarity & 0.358 & -- \\
         & Contr. Diversity * Familiarity & 0.187 & -- \\
         & Contr. Balance * Familiarity & 1.0 & -- \\
         & Contr. Diversity * Contr. Balance * Familiarity & 0.38 & -- \\
       \bottomrule
    \end{tabular}
     }
\end{table}

\section{Data Distribution}
\label{app:data_distribution}

\begin{figure*}[tp]
    \centering
    \includegraphics[width=1.0\textwidth]{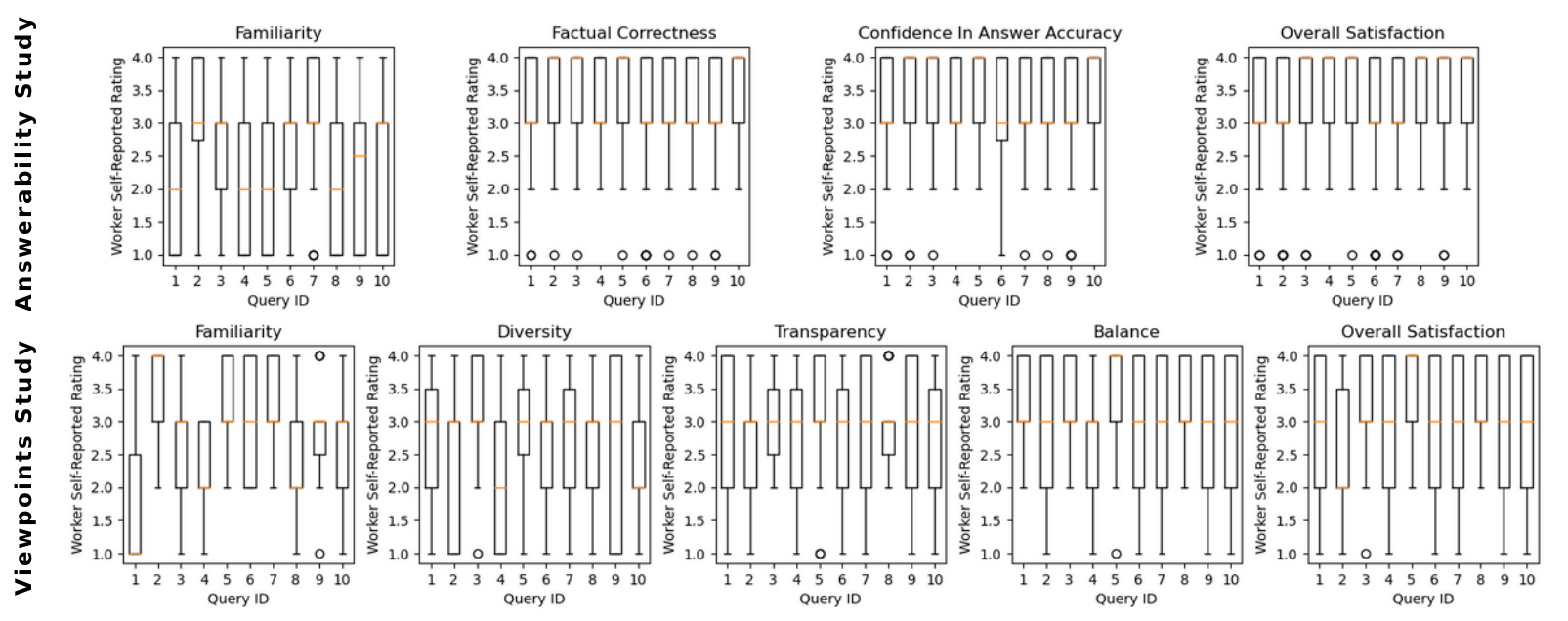}
    \caption{Distribution of user-judged response dimensions per query in the two user studies. 
    }
    \label{fig:scores_distribution}
\end{figure*}

Figure~\ref{fig:scores_distribution} presents the distribution of ratings for user-judged response dimensions, topic familiarity, and overall satisfaction per query for both user studies.

\section{Qualitative Analysis}
\label{app:qualitative_analysis}

To validate our findings, we characterize workers' user experience by analyzing their natural language comments.
We manually inspect all the 960 worker comments in the \studyone{} and 270 in the \studytwo{}. 
We followed an inductive approach~\citep{Williams:2008:} to identify themes in the comments. After consensus among the authors, one of the authors labeled all comments.
Next, we counted how many workers mentioned a particular aspect.\footnote{While adhering to an established qualitative analysis approach, the authors acknowledge that personal interpretation may introduce some degree of subjectivity in the interpretation and categorization of the data.}
In the reporting below, numbers in parentheses indicate the proportion of high (when the aspect mentioned in the comments is positive) or low (when the aspect is negative) satisfaction ratings corresponding to specific aspects mentioned in the comments. For instance, if three workers mention a positive aspect like ``factual correctness'' in the open text field, but only two assign high satisfaction scores on the four-point Likert scale, this is reported as 2/3. Conversely, if a negative aspect like ``bias'' is identified by five workers and four of them express low satisfaction, it is recorded as 4/5. 

Coherence, fluency, naturalness, details, and logic of the response mentioned in the comments are almost always accompanied by high satisfaction ratings (178/185 in the \studyone{} and 23/23 in the \studytwo{}).
In the \studyone{}, comments mentioning positive aspects such as factual correctness (126/133), information completeness (99/100), agreement with the response (59/60), presence (53/64), and credibility (18/23) of the source are accompanied by high satisfaction ratings (3 or 4 on the Likert scale). However, high satisfaction ratings are not always paired with positive comments. Some comments associated with high satisfaction ratings indicate negative aspects, such as lack of source (4/21) or invalid source (2/11). Additionally, highlighting missing or incomplete information (60/156) does not always cause a decrease in the satisfaction rating. 
In the \studytwo{}, positive comments indicating high diversity (55/58), balance (6/6), lack of bias (12/13), completeness of the provided response (22/22), or agreement with the answer (8/9) are accompanied by high satisfaction ratings. However, some responses describing negative aspects such as bias (14/25) or lack of diversity (43/64) are still given high satisfaction ratings. Most of the responses described as not diverse (43/64) or imbalanced (12/22) are accompanied by low satisfaction ratings (1 or 2 on the Likert scale).
Additional aspects mentioned in the comments include the usefulness (33/35 comments accompanied by high satisfaction rating) and subjectivity (10/24 comments accompanied by low satisfaction rating) of the response in the \studyone{}, and lack of source (4/12 comments accompanied by low satisfaction rating) in the \studytwo{}. It is worth pointing out that while usefulness is a common indicator of successful completion of the search task~\citep{Cambazoglu:2021:CHIIR, Liu:2023:CHIIR}, it is only mentioned in 2.8\% of the comments. 

Although satisfaction ratings are skewed (see~Figure~\ref{fig:scores_distribution}) 
and other scales such as magnitude estimation~\cite{Turpin:2015:SIGIR} may give us more informative ratings, they are roughly aligned with the dimensions we aim to capture. It implies that the dimensions we use to differentiate between response variants impact user satisfaction. 
Comments from the \studyone{} suggest that satisfaction is associated with both factual correctness and source validity. The frequent user references to factual correctness in comments imply a significant focus on this aspect when evaluating responses. Even though we observe a high correlation between user-reported response dimensions and overall satisfaction, we do not observe a statistically significant effect of the controlled factual correctness on user ratings for this response dimension. This implies that users find these response dimensions important and associate them with their satisfaction, but they are not able to identify factuality correctly in the responses.
Additionally, one-way ANOVA for the \studyone{} revealed that the overall satisfaction is affected by the query and topic familiarity, not the controlled response dimensions. This also explains why the response dimensions mentioned in the comments do not completely align with the actual flaws in the responses.
The results' sensitivity to the topics may suggest that including more queries in a further studies might reveal the effect of factual correctness and source validity on overall satisfaction.
In the \studytwo{}, our qualitative analysis shows that user satisfaction is more linked to viewpoint diversity and response completeness than information balance, differing from quantitative findings. It can follow from the fact that the concept of response diversity is better understood by users and is easier to identify. 
Nevertheless, the qualitative analysis shows that selected response dimensions are indeed common indicators of user satisfaction.

\section{Lessons Learned from Executing User Studies}
\label{app:lessons}

Reflecting on our experiences from conducting these user studies, several key lessons that may be found useful by the community have become apparent.
\begin{enumerate}
    \item  The effectiveness of these studies depends on the careful selection of representative queries and responses to the problems being investigated. 
    \item Incorporating validation steps, especially in experiments that involve subjectivity, helps mitigate biases introduced by study designers. 
    \item Implementing attentiveness checks is crucial for ensuring the quality of collected data and maintaining the credibility of the gathered information. 
    \item While qualitatively analyzing responses from crowd workers in natural language may incur higher costs, it can reveal unforeseen dimensions and challenges. Furthermore, natural language responses for open-ended questions serve as reliable indicators of data quality—fluent, relevant, and informative responses from crowd workers typically accompany meaningful data.
    \item Developing operational definitions of explored dimensions and refining them iteratively during the initial stages of experimentation and design fosters a deeper understanding of the dimensions under examination and facilitates necessary adjustments before data collection begins.
\end{enumerate} 

\end{document}